\documentclass[doublecol]{epl2}

\title{Electrostatic control of quantum dot entanglement\\ induced by
  coupling to external reservoirs} \shorttitle{Electrostatic control
  of quantum dot
  entanglement} 

\author{E. del Valle, F.P. Laussy and C. Tejedor}
\shortauthor{E. del Valle \etal}

\institute{Departamento de F\'{\i}sica Te\'orica de la Materia
Condensada, Universidad  Aut\'onoma de Madrid, Cantoblanco 28049
Madrid, Spain}


\abstract{We propose a quantum transport experiment to prepare and
  measure charge-entanglement between two electrostatically defined
  quantum dots. Coherent population trapping, as realized in cavity
  quantum electrodynamics, can be carried out by using a third quantum
  dot to play the role of the optical cavity. In our proposal, a
  pumping which is quantum mechanically indistinguishable for the
  quantum dots drives the system into a state with a high degree of
  entanglement. The whole effect can be switched on and off by means
  of a gate potential allowing both state preparation and entanglement
  detection by simply measuring the total current.}

\begin{document}

\maketitle

\section{Introduction}
Experiments on transport through quantum dots (QDs) have recently
experienced such a development that it is now possible to reproduce
many of the phenomena that the field of quantum optics, involving
atoms, has been exploring for many years~\cite{yamamoto,brandes05}.
In particular, state preparation and manipulation of one or more QDs
is nowadays feasible by controllable external means such as gate and
bias potentials plus either continuous or AC electric and/or
magnetic fields~\cite{note1}. Each QD can be considered as a
\emph{qubit} with the lower state~$|0\rangle$ corresponding to a
neutral QD and the upper state~$|1\rangle$ to having one extra
electron in the QD. Due to Coulomb blockade, charging the QD with
more than one electron requires an energy that can be considered
infinite for any practical purpose, reducing the Hilbert space to
that spanned by the two mentioned states. A key point for building
up logical gates is the capability of producing entanglement between
two \emph{qubits}. Although entanglement effects in transport
through double QDs have been extensively
studied~\cite{hayashi03,note1,vorrath03,marquardt03,brandes05,michaelis06,diasdasilva06,lambert07},
quantum optics's \emph{know-how} can be helpful to control such a
quantum feature.

The aim of this work is to make a proposal for experimentally
preparing and measuring a charge-entangled state of two QDs. We
start from the pioneering ideas of ref.~\cite{michaelis06}, where
the constraint of having no more than one electron in the whole
system allows the population trapping in a dark-entangled state. The
same configuration does not give the desired results within a regime
more experimentally accessible (more than one electron in total) as
we will show below. We obtain interesting results (in experimentally
accessible conditions) when adding a new and crucial physical effect
borrowed from cavity quantum electrodynamics (CQED): cross-terms in
the incoherent pumping~\cite{perea05,troiani06b} of two QDs in a
microcavity can produce entanglement by driving the system to a
\emph{quasi-dark} state (a state which is only weakly coupled to the
cavity)~\cite{delvalle07}. However, it is difficult to detect this
process by optical means. In the framework of transport, the role of
the cavity can be played by a third QD and both incoherent pumping
of the QDs and cavity photon emission find their counterpart for the
transport realization in the tunneling processes produced by the
application of a bias. We will show here that, apart from all the
analogies, there is an important advantage in doing transport:
entanglement could be easily detected in the same setup that
prepares the equivalent to the quasi-dark (entangled) state.

\section{The system and its dynamics}
\begin{figure}[!hbt]
\centering
\includegraphics[width=.9\linewidth]{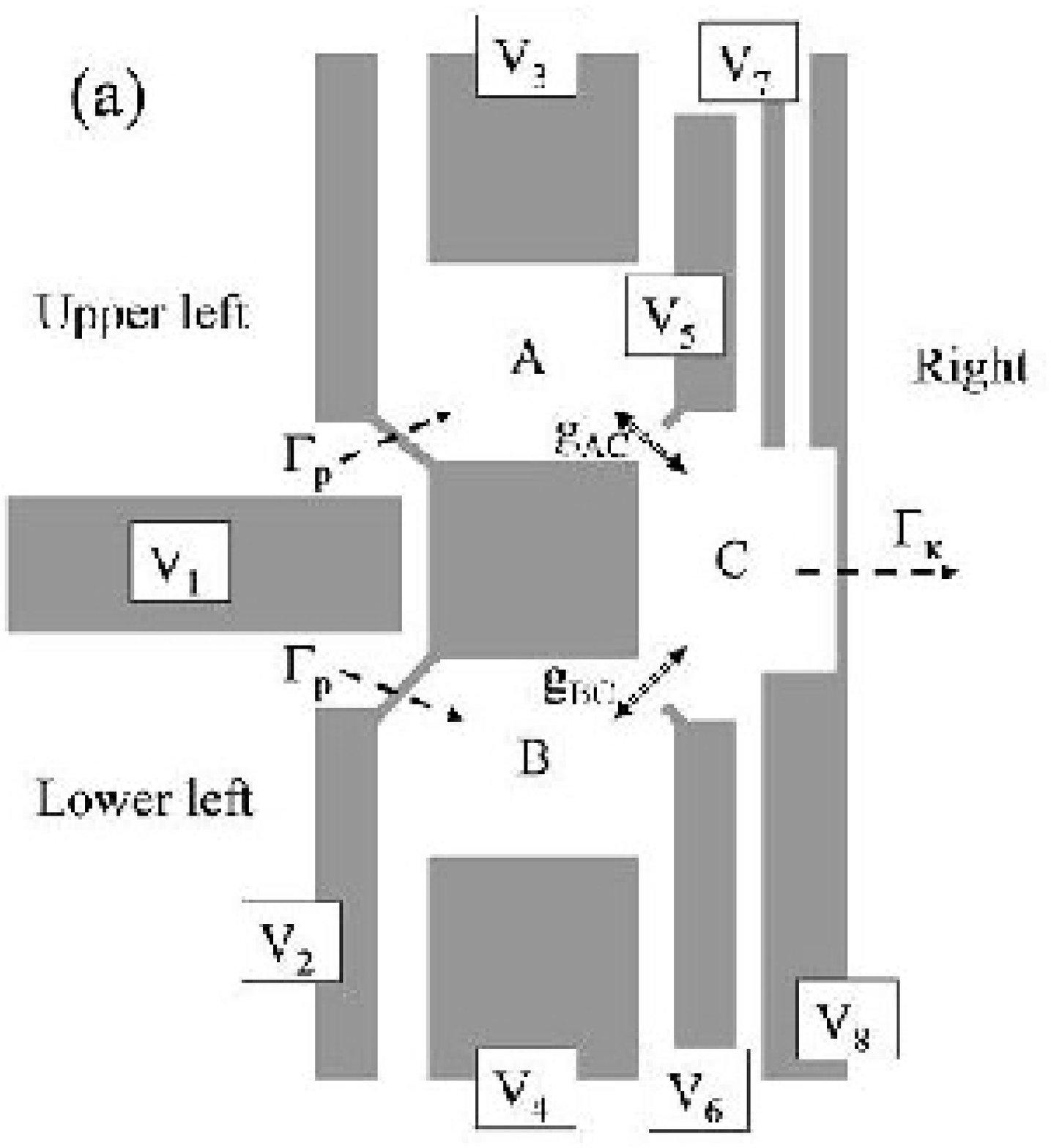}
\includegraphics[width=.9\linewidth]{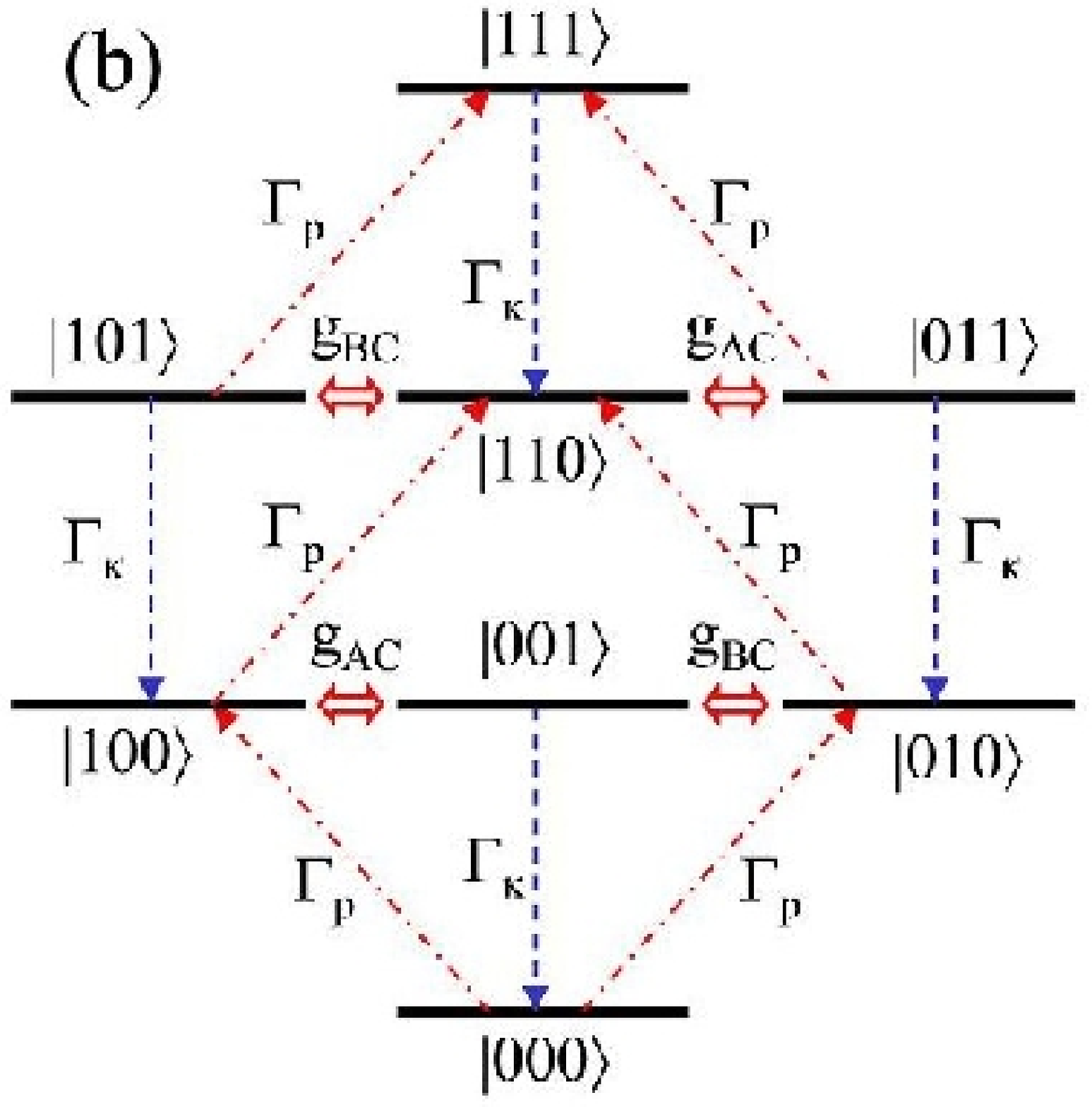}
\caption{Color online. (a) Scheme of the proposed setup, with a
  two-dimensional electron gas depleted by 8 gate potentials $V_i$.
  $V_3$, $V_4$ and $V_7$ control the levels of the QDs. $V_2$ and
  $V_8$ control $\Gamma_\mathrm{p}$ and $\Gamma_\kappa$. $V_5$ and
  $V_6$ control $g_{AC}$ and $g_{BC}$. Switching $V_1$ from ``on'' to
  ``off'' tunes the pumping of $A$ and $B$ from distinguishable to
  indistinguishable quantum mechanically. The current is induced by a
  bias from left to right. (b) Scheme of the dynamics in the Hilbert
  space spanned by $|n_A,n_B,n_C\rangle$. We simplify the plot by
  setting detunings to zero.}
\label{fig:1}
\end{figure}

Our system is presented in figure~\ref{fig:1}(a). A two-dimensional
electron gas is depleted in some regions by means of a series of
gate potentials. A bias applied from the left to the right lead
produces two tunneling processes: incoherent population of QDs~$A$
and $B$ as well as electron current from QD~$C$ to the right lead.
QDs $A$ and $B$ are coherently coupled to QD~$C$ (acting as the
cavity in CQED). The gate-potentials $V_3$, $V_4$ and $V_7$ are
designed to control the levels of the three \emph{qubits}, $V_2$ and
$V_8$ control the in- ($\Gamma_\mathrm{p}$) and out-
($\Gamma_\kappa$) tunneling rates while the gates $V_5$ and $V_6$
control the coherent couplings $g_{AC}$ and $g_{BC}$ respectively.
The crucial novelty with respect to that of~\cite{michaelis06} is
the gate $V_1$.  Switching on and off the $V_1$ gate, one can
experimentally tune from a quantum mechanically distinguishable
($V_1$ on) to an indistinguishable ($V_1$ off) pumping of the two
QDs $A$ and $B$. Let us first see how this appears in the quantum
mechanical description of the transport through the whole system and
later what are the physical effects that can be tuned in this setup.

Coulomb blockade on each QD limits the Hilbert space to that spanned
by a basis of 8 states $|n_A,n_B,n_C\rangle$ (depicted in
figure~\ref{fig:1}(b)) with $n_{A,B,C}=0,1$. Cross term effects that
we are going to concentrate on only occur for electrons with the
same spin. We therefore consider the system under the action of an
in-plane magnetic field and neglect the spin. Both intra-dot Coulomb
blockade and spin polarisation stand within an experimentally
accessible regime. To reduce the Hilbert space to the lowest four
states in figure~\ref{fig:1}(b), as is done in~\cite{michaelis06},
would require an extremely high inter-dot Coulomb repulsion,
something unreasonable in a system such as the one shown in
figure~\ref{fig:1}(a). The coherent part of the dynamics is
controlled by the Hamiltonian (we take $\hbar =1$):
\begin{equation}
\label{hamilt} H_0=\sum _{i=A,B} \Delta _i c^\dagger_i c_i+ g_{iC}
(c^\dagger_i c_C+c^\dagger_C c_i)
\end{equation}
where $c_i$, $c_i^\dagger$ are annihilation and creation operators
of an electron in QD~$i$. We have taken the level of the QD~$C$ as
the origin of energies so that only the detunings $\Delta_A$,
$\Delta_B$ and the couplings $g_{AC}$, $g_{BC}$ are relevant.

Following the methods of quantum
optics~\cite{walls,yamamoto,brandes05}, we obtain a master equation
that describes the whole dynamics of the density matrix~$\rho$
within the Born-Markov
approximation~\cite{ficek02,perea05,walls,troiani06b,delvalle07}:
\begin{eqnarray}
\label{mastereq}
\frac{d \rho}{dt}&=i \lbrack \rho, H_0 \rbrack + ig_{AB} \lbrack \rho ,
(c^\dagger_A c_B+ c^\dagger_B c_A) \rbrack \nonumber \\
&+\frac{\Gamma_\kappa}{2}
\big(2 c_C \rho c_C^\dagger -c_C^\dagger c_C \rho-\rho c_C^\dagger c_C \big) \nonumber \\
&+\frac{\gamma_{d}}{2}\sum_{i=1}^8 \big(2\mathcal{P}_{i} \rho
\mathcal{P}_{i}-
\mathcal{P}_{i} \rho-\rho \mathcal{P}_{i} \big) \nonumber \\
+\sum_{i,j=A,B}&
\frac{\Gamma_\mathrm{p}\delta_{ij}+\Gamma_{AB}(1-\delta_{ij})}{2}\big(2c_i^\dagger
\rho c_j-c_j c_i^\dagger \rho-\rho c_j c_i^\dagger\big) \nonumber \\
\label{mastereq}
\end{eqnarray}
where~$\mathcal{P}_{i}=|n_A,n_B,n_C\rangle\langle n_A,n_B,n_C|$.  As
is well known, the equation of motion consists of a coherent part and
a dissipative part.  The coherent part has the contribution of the
Hamiltonian dynamics $H_0$ on the first line while the incoherent part
features the in- ($\Gamma_\mathrm{p}$) and out- ($\Gamma_\kappa$)
tunneling processes (on lines 2 ~\& 4), and dephasing at rate
$\gamma_\mathrm{d}$ (line 3). There are also two extra ingredients
which appear in eq.~(\ref{mastereq}) from considering couplings to
reservoirs up to the second
order~\cite{walls,ficek02,perea05,troiani06b,delvalle07}. These two
new terms are responsible for the new physics that appears in this
problem.

The first term is an incoherent contribution to the dynamics given
by a cross Lindblad term in the pumping part, on line 4, involving
operators with different QD indexes. It appears when the QDs~$A$ and
$B$ are pumped from the same reservoir (left lead) in a complete
indistinguishable (quantum) way.  In such a case, the corresponding
rate of pumping is given by $\Gamma_{AB}=\Gamma_\mathrm{p}$. This
corresponds to switching off the gate $V_1$. By smoothly switching
it on, the upper and lower parts of the left lead separate from each
other. $\Gamma_{AB}$ is reduced down to the situation in which the
two left reservoirs become completely independent and
$\Gamma_{AB}=0$. Cross terms in the pumping are therefore
experimentally controllable.

The second term on the right hand side of line 1, is a
generalization of that contributing to the Lamb-shift of a single
two-level system~\cite{walls,ficek02,perea05,troiani06b,delvalle07}.
So, a common indistinguishable pumping provides an effective
F\"{o}rster-like coupling between QDs $A$ and $B$ with coupling
parameter $g_{AB}=2\Gamma_{AB}$~\cite{ficek02}. This coupling, which
is thus coherent, contributes to enhancing the entanglement between
QDs $A$ and $B$, although it stems from the incoherent pumping. Once
again, one can experimentally control this mechanism from switching
on $V_1$, which gives $g_{AB}=0$, to switching it off, which gives
$g_{AB}=2\Gamma_\mathrm{p}$.

Once the dynamics of the density matrix is known, one can compute
the current flowing through the system, which is an easily measured
experimental quantity. We use the input-output formulation of
optical cavities~\cite{walls}, in which the role of the cavity is
played by the QD~$C$. The current is simply given by
$I=\Gamma_\kappa\langle {c_C}^\dagger c_C\rangle$. In the stationary
limit, which is the case of greatest interest to us, the master
equation~(\ref{mastereq}) simplifies to a set of~64 linear
equations, plus the normalization condition $\mathrm{Tr}(\rho)=1$.
In order to correlate the current with the entanglement of $A$ and
$B$, we quantify the latter by the tangle~\cite{wootters98}, that in
our case reads
\begin{equation}
  \label{eq:WedJul4202452BRT2007}
  \tau = [ \max \{0, 2(|\widetilde{\rho}_{10,01}|-\sqrt{\widetilde{\rho}_{00,00}\widetilde{\rho}_{11,11}})\}]^2
\end{equation}
in terms of the matrix elements of
$\widetilde{\rho}=\mathrm{Tr}_C(\rho)$~\cite{delvalle07}.
\begin{figure}[!hbt]
\centering
\includegraphics[width=\linewidth]{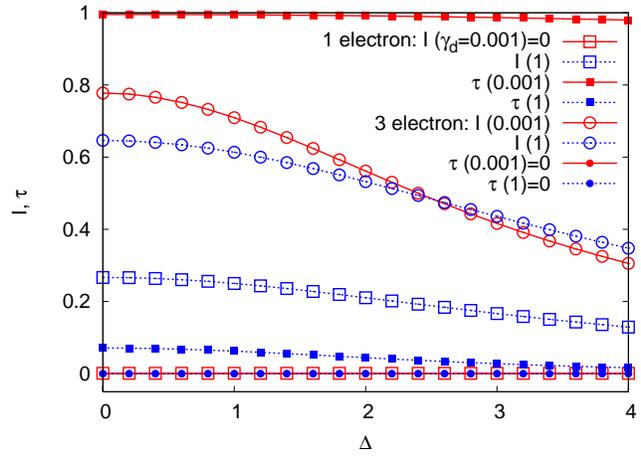}
\caption{Color online. Current intensity $I$ (empty symbols) and
  tangle $\tau$ (full symbols) as a function of detuning $\Delta$ when
  $V_1$ is so large that pumpings to $A$ and $B$ are distinguishable
  from each other, i.e. $\gamma_{AB}=0$. Two different dephasings are
  considered: $\gamma_\mathrm{d}=0.001$ (in solid-red) and
  $\gamma_\mathrm{d}=1$ (in dashed-blue).
  $\Gamma_\mathrm{p}=\Gamma_\kappa=2$ and $g_{BC}=1$. Energies and
  rates in units of $g_{AC}=1$.  Strong Coulomb blockade is considered
  either inter-QD (maximum charge of~1 electron in the whole system,
  plotted with squares) or just intra-QD (maximum charge of 1+1+1=3
  electrons, plotted with circles). Red and blue full circles coincide
  to zero (as the tangle in the three-electron case for all dephasing
  rates is zero) and therefore they appear superimposed in the plot.
  Also the current in the case of~1 electron and negligible dephasing
  ($\gamma_\mathrm{d}=0.001$) is zero.}
\label{fig:2}
\end{figure}

\section{Results}
Let us discuss the results to be expected from the setup we propose.
We consider that the QDs $A$ and $B$ are equal, $\Delta_A
=\Delta_B=\Delta$, what can be achieved by adjusting independently the
gates $V_3$ and $V_4$. The entanglement can be manipulated by having
different couplings $g_{AC}$ and $g_{BC}$ as controlled by the gates
$V_5$ and $V_6$. Hereafter, all the couplings and rates will be given
in units of $g_{AC}=1$.  

First of all, we analyze the adequacy of truncating the Hilbert space
basis to only four states~\cite{michaelis06}. For this purpose we
consider the simple case of neglecting cross-terms and estimate the
total mean number of excitations in the system. For a system that is
pumped with a total rate $P_{tot}$, decays with $\kappa_{tot}$ and
that has a saturation limit $S$, we find that the mean total number of
excitations in the strong coupling regime is given by the expression:
\begin{equation}
  \label{eq:mean_number}
  \langle n \rangle= S \frac{P_{tot}}{P_{tot}+\kappa_{tot}}
\end{equation}
In our case where~$P_{tot}=2\Gamma_p$ (two QDs each pumped at
rate~$\Gamma_p$), $\kappa_{tot}=\Gamma_\kappa$ and~$S=3$ (maximum of
three electrons in the system), the general formula
eq.~(\ref{eq:mean_number}) gives, in the symmetric case that we
consider in this paper, $\Gamma_{P}=\Gamma_{\kappa}=\Gamma$:
\begin{equation}
  \label{eq:mean_number_system}
  \langle n \rangle= 3 \frac{2 \Gamma}{2\Gamma+\Gamma}=2
\end{equation}
This result, that agrees with numerical calculations, is the first
indication of the inadequacy of truncating the Hilbert space to only
one electron.  Moreover, we find that the relevant magnitudes under
study ($I$ and $\tau$) depend strongly on the truncation.
Figure~\ref{fig:2} shows $I$ and $\tau$ as a function of the detuning
for two different dephasing rates, $\gamma_\mathrm{d}=1$ and
$\gamma_\mathrm{d}=0.001$, both with truncation (a maximum of one
electron in the system) and without truncation (a maximum of three
electrons in the system). When truncation is not imposed, $\tau$ is
always zero so that entanglement is not achieved.

The approximation of keeping just one electron in the whole system
forces the steady state of QDs $A$ and $B$ to be a singlet
$|\mathrm{S~0}\rangle=(|100\rangle - |010\rangle)/\sqrt{2}$ (with QD
$C$ in the vacuum). This implies a tangle of one and no current
passing through the system if the dephasing is
negligible~\cite{michaelis06}. This is an example of a trapping
mechanism. The pumping is populating both the singlet and its
symmetric counterpart, the triplet state
$|\mathrm{T~0}\rangle=(|100\rangle + |010\rangle)/\sqrt{2}$.  However,
when the couplings are equal $g_{AC}=g_{BC}$, the singlet is dark,
does not couple to other states and finally stores all the excitation
of the system in the steady state. Therefore, when more than one
electron is allowed, this trapping mechanism breaks as also the states
$|11n_C\rangle$ become pumped. In the absence of cross terms, the
tangle drops to zero and there is current through the system. A
negative result to be drawn from figure~\ref{fig:2} is that without
cross terms, in the actual case of more than one electron, there is no
entanglement to be expected experimentally.
\begin{figure}[!hbt]
\centering
\includegraphics[width=\linewidth]{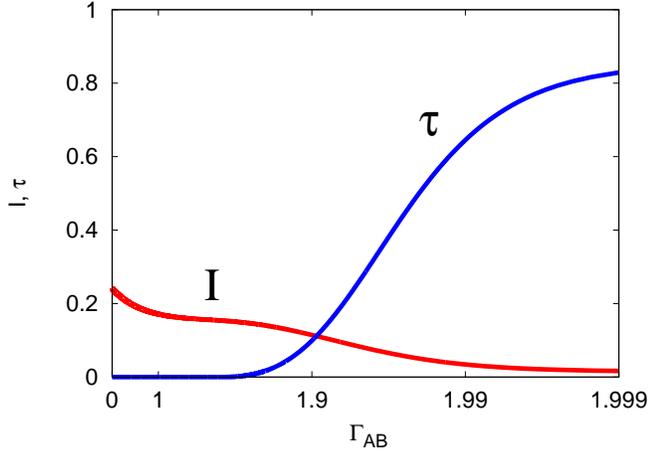}
\caption{Color online. Current intensity $I$ (in red) and tangle
  $\tau$ (in blue) as a function of $\Gamma_{AB}$ (in logarithmic
  scale).  This is experimentally controlled by varying $V_1$ from a
  large value (giving $\Gamma_{AB}=0$) to zero (giving
  $\Gamma_{AB}=\Gamma_\mathrm{p}$). $\Delta =4$,
  $\Gamma_\mathrm{p}=\Gamma_\kappa=2$, $\gamma_\mathrm{d}=0.001$ and
  $g_{BC}=0.7$ with energies and rates in units of $g_{AC}=1$. Fast
  rising of $\tau $, detectable by fast quenching of $I$, is due to
  the switching on of a quantum mechanically indistinguishable
  pumping.  Only intra-QD Coulomb blockade is considered (maximum
  charge of 3 electrons).}
\label{fig:3}
\end{figure}

Our main finding is the entanglement induced by cross terms in the
dynamics, enhanced by the coherent coupling between $A$ and $B$.
Hereafter we consider the general case, i.e., without truncation to
only one electron. Figure~\ref{fig:3} shows $I$ and $\tau$ as a
function of $\Gamma_{AB}$ for the larger detuning $\Delta=4$ and the
lowest dephasing $\gamma_\mathrm{d}=0.001$ considered in
figure~\ref{fig:2}.  An important fact is that now the couplings
$g_{AC}$, $g_{BC}$ must be slightly different (for instance
$g_{BC}=0.7$) so that the singlet is not completely dark, but a
quasi-dark state weakly coupled to the rest of the system (with a
coupling given by $|g_{AC}-g_{BC}|/\sqrt{2}$). When the gate $V_1$ is
completely switched on, $\Gamma_{AB}=0$ and, as it happened in
figure~\ref{fig:2}, there is current larger than $I=0.2$, implying no
entanglement. Increasing $\Gamma_{AB}$ by quenching the gate $V_1$
does not affect the behavior of the system until the regime where
cross terms apply fully is reached. Here, when $\Gamma_{AB}$ tends to
$\Gamma_\mathrm{p}$, adding cross terms in eq.~\ref{mastereq}
translates in pumping only the symmetric states (under QDs $A$, $B$
exchange). Therefore the incoherent pump with cross terms neither
excites the singlet nor induces decoherence of it. This fact, together
with the weak link between the singlet state and the other levels,
results in a slow coherent transfer of population to the singlet
$|\mathrm{S~0}\rangle$, which can be described as a quasi-dark state
free of decoherence. This novel trapping mechanism is enhanced
strongly by the direct coupling $g_{AB}$, also induced by the cross
pump. In this case, the tangle becomes close to its highest possible
value of~1. The detectable manifestation is a sharp reduction of the
current through the system, as QD $C$ is practically empty.  This
means a clear way of entangled state preparation between QDs $A$ and
$B$ as well as a straightforward measurement associated to its
occurrence (drop of the current).

\begin{figure}[!hbt]
\centering
\includegraphics[width=0.9\linewidth]{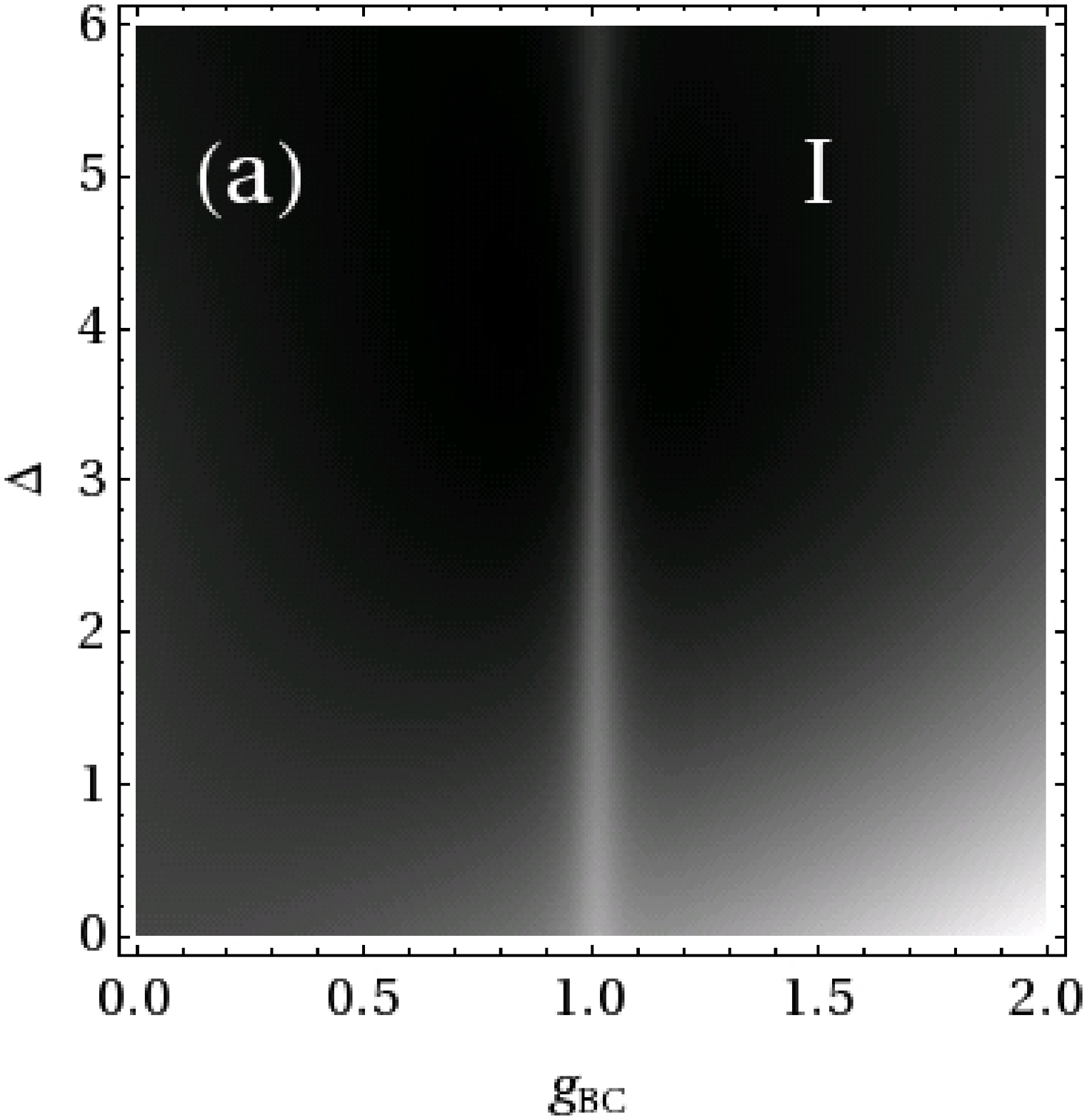}
\includegraphics[width=0.9\linewidth]{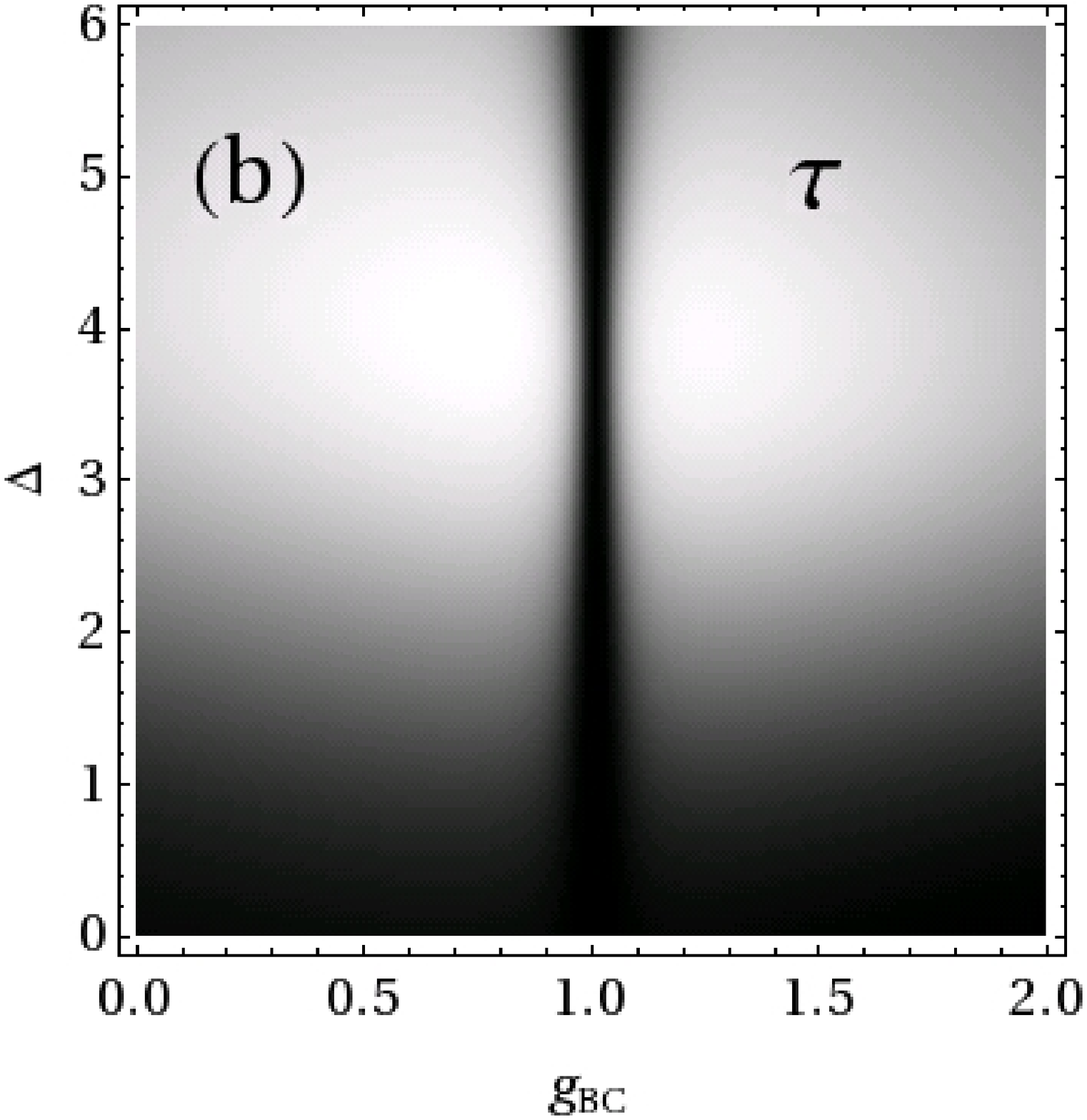}
\caption{Current intensity $I$~(a) and tangle $\tau$~(b)
  in density plots as a function of $g_{BC}$ and $\Delta$ in the
  quantum mechanically indistinguishable case
  $\Gamma_{AB}=\Gamma_\mathrm{p}$.
  $\Gamma_\mathrm{p}=\Gamma_\kappa=2$ and $\gamma_\mathrm{d}=0.001$.
  Energies and rates are in units of $g_{AC}=1$. Bright areas
  correspond to maximum values of current and tangle and the dark ones
  to zero.}
\label{fig:4}
\end{figure}

Finally, we want to show how entanglement induced by cross terms
depends on the coherent part of the dynamics controlled by $H_0$.
For this purpose, figure~\ref{fig:4} presents current~$I$~(a) and
tangle $\tau$~(b) as a function of the detuning $\Delta$ and the
coherent coupling $g_{BC}$ (always in units of $g_{AC}$). The tangle
plot shows that detuning is needed to generate a high degree of
entanglement. As we explained, also slightly different couplings
$g_{AC}$ and $g_{BC}$ are necessary to create the quasi-dark state.
In figure~\ref{fig:3} we were giving results for the situation with
highest tangle ($\tau=0.85$), that is $g_{BC}=0.7$ and $\Delta=4$,
corresponding also to lowest current ($I=0.01$). On the other hand,
for the symmetric case $g_{AC}=g_{BC}$, the singlet is completely
dark and therefore there is no entanglement, as we also showed in
figure~\ref{fig:2}. In this case the current is nonzero. The
correlation between high tangle and negligible current and vice
versa is clear from figure~\ref{fig:4}.

\section{Conclusions}
In conclusion, we present a proposal of quantum transport experiment
for preparing and measuring in the steady state a charge-entangled
state of two non-interacting QDs: entanglement is produced by means
of a quantum mechanically indistinguishable pumping when each QD is
coherently coupled to a third one, playing the role of a cavity in
CQED. Indistinguishable pumping produces cross Lindblad terms and a
coherent direct coupling between the dots. Their combined effect
results in a coherent trapping mechanism in an entangled state. This
source of entanglement can be switched on and off by means of a gate
potential.  This allows both state preparation and entanglement
detection by simply measuring the total current.

\acknowledgments This work has been partly supported by the Spanish
MEC under contracts Consolider-Ingenio2010 CSD2006-0019,
MAT2005-01388, NAN2004-09109-C04-4, and by CAM under Contract
S-0505/ESP-0200.

\end{document}